\documentclass[11pt]{article}

\usepackage[final]{acl}

\usepackage{times}
\usepackage{latexsym}
\usepackage[T1]{fontenc}
\usepackage[utf8]{inputenc}
\usepackage{microtype}
\usepackage{inconsolata}
\usepackage{graphicx}
\usepackage{booktabs}
\usepackage{multirow}
\usepackage{tabularx}
\usepackage{fvextra}
\usepackage{amssymb}
\DefineVerbatimEnvironment{SPARQL}{Verbatim}{breaklines=true,breakanywhere=true,fontsize=\small}

\title{PIPE-RDF: An LLM-Assisted Pipeline for Enterprise RDF Benchmarking}

\author{
  Suraj Ranganath \\
  UC San Deigo \\
  \texttt{suranganath@ucsd}
}

\begin{document}
\maketitle

\begin{abstract}
Enterprises rely on RDF knowledge graphs and SPARQL to expose operational data through natural language interfaces, yet public KGQA benchmarks do not reflect proprietary schemas, prefixes, or query distributions. We present PIPE-RDF, a three-phase pipeline that constructs schema-specific NL--SPARQL benchmarks using reverse querying, category-balanced template generation, retrieval-augmented prompting, deduplication, and execution-based validation with repair. We instantiate PIPE-RDF on a fixed-schema company-location slice (5\,000 companies) derived from public RDF data and generate a balanced benchmark of 450 question--SPARQL pairs across nine categories. The pipeline achieves 100\% parse and execution validity after repair, with pre-repair validity rates of 96.5\%--100\% across phases. We report entity diversity metrics, template coverage analysis, and cost breakdowns to support deployment planning. We release structured artifacts (CSV/JSONL, logs, figures) and operational metrics to support model evaluation and system planning in real-world settings. Code is available at \url{https://github.com/suraj-ranganath/PIPE-RDF}.
\end{abstract}

\section{Introduction}
Enterprise teams increasingly rely on RDF knowledge graphs to represent relational data and use SPARQL to retrieve answers for analysts, operations teams, and customer-facing applications. Modern LLMs, from BERT and T5 to GPT-3 and LLaMA, lower the barrier for translating natural language into formal queries \citep{devlin2019bertpretrainingdeepbidirectional,raffel2023exploringlimitstransferlearning,brown2020languagemodelsfewshotlearners,touvron2023llamaopenefficientfoundation}. In practice, model performance is tightly coupled to the target schema, namespace conventions, and query distribution. Public KGQA benchmarks (e.g., LC-QuAD, QALD-9-plus, Mintaka) are built on DBpedia or Wikidata and do not capture enterprise-specific ontologies or internal vocabulary \citep{Trivedi_2017,perevalov2022qald9plusmultilingualdatasetquestion,sen2022mintakacomplexnaturalmultilingual}. As a result, teams evaluating prompt strategies, retrieval augmentation, or fine-tuning (e.g., QLoRA) lack reliable in-domain benchmarks \citep{dettmers2023qloraefficientfinetuningquantized}.

PIPE-RDF addresses this gap with a pipeline for constructing proprietary, category-balanced NL--SPARQL benchmarks. The pipeline emphasizes (i) grounding question templates in graph facts via reverse querying, (ii) balancing complex query categories to avoid evaluation bias, (iii) retrieval-augmented prompting for structural alignment, and (iv) execution-based validation and repair. These requirements mirror the needs of industry teams who must iterate quickly, monitor quality, and maintain benchmarks as schemas evolve. We position PIPE-RDF as an \emph{emerging} industry system: a deployable pipeline that bridges research methodology with production constraints (latency, maintenance, and data governance) without requiring proprietary model training or disclosure.

\textbf{Contributions.} We:
\begin{itemize}
  \item propose a three-phase benchmark generation pipeline tailored to enterprise RDF graphs, with reverse querying and execution validation;
  \item implement PIPE-RDF on a fixed-schema company-location slice (Schema C) and release balanced benchmark artifacts;
  \item report operational metrics (parse validity, execution success, empty results, latency, complexity, pre-repair rates, entity diversity) from a final 450-pair run;
  \item provide guidance for maintaining and scaling organization-specific benchmarks.
\end{itemize}

\section{Background and Related Work}
\textbf{Benchmark coverage and query taxonomies.} KGQA benchmarks often over-represent simple factoid queries. Mintaka explicitly balances query categories (counting, comparative, superlative, ordinal, multi-hop, intersection, difference/negation, yes/no, and generic) to avoid skewed evaluation \citep{sen2022mintakacomplexnaturalmultilingual}. Public datasets such as LC-QuAD and QALD-9-plus provide valuable coverage but remain tied to public graphs and fixed schemas \citep{Trivedi_2017,perevalov2022qald9plusmultilingualdatasetquestion}. Dynamic benchmarking efforts such as Dynabench further highlight the need for iterative, real-world evaluation \citep{kiela2021dynabenchrethinkingbenchmarkingnlp}.

\textbf{Retrieval augmentation and prompting.} Retrieval-augmented generation (RAG) improves text-to-query performance by conditioning on relevant examples and schema cues \citep{lewis2021retrievalaugmentedgenerationknowledgeintensivenlp}. In practice, RAG offers a low-cost alternative to fine-tuning when schema-specific examples are scarce. PIPE-RDF maintains category-specific retrieval banks to provide structurally similar NL--SPARQL pairs during generation, reducing cross-category drift and improving template grounding.

\textbf{Validity and evaluation.} Grammar-constrained decoding (e.g., PICARD) improves syntactic correctness for structured query generation \citep{Scholak_2021}. SPARQL-specific metrics such as SP-BLEU/SP-F1 normalize variables for fairer matching \citep{Rony_2022}, while code-oriented structural metrics such as CodeBLEU inspire future extensions \citep{ren2020codebleumethodautomaticevaluation}. PIPE-RDF focuses on execution, parsing, and structural complexity metrics that are directly actionable for deployment and troubleshooting.

\textbf{LLM-driven workload generation for analytics.} Recent work on cross-model analytics workload generation uses a two-stage design (reverse query generation followed by NL query generation), explicit schema/strategy controls, and detailed error analyses across model families \citep{agarwal2024crossmodel}. We adopt the same design principle of schema-grounded workload construction, and we report strategy-level coverage and failure modes to strengthen enterprise readiness.

\textbf{Comparison with existing approaches.} Table~\ref{tab:comparison} contrasts PIPE-RDF with prominent KGQA benchmark generation methodologies. Unlike static public benchmarks, PIPE-RDF introduces reverse querying to ground templates in graph facts, category-aware retrieval banks to prevent structural drift, and execution-based repair loops for automated quality control. These features address enterprise requirements for schema fidelity and operational observability that public benchmarks cannot provide.

\begin{table}[t]
  \centering
  \small
  \resizebox{\linewidth}{!}{
  \begin{tabular}{lcccc}
    \toprule
    Feature & LC-QuAD & QALD & Mintaka & \textbf{PIPE-RDF} \\
    \midrule
    Schema fidelity & Public & Public & Public & \textbf{Fixed} \\
    Reverse querying & -- & -- & -- & \textbf{$\checkmark$} \\
    Category balance & Partial & -- & \textbf{$\checkmark$} & \textbf{$\checkmark$} \\
    Repair loop & -- & -- & -- & \textbf{$\checkmark$} \\
    Operational metrics & -- & -- & -- & \textbf{$\checkmark$} \\
    RAG retrieval & -- & -- & -- & \textbf{$\checkmark$} \\
    \bottomrule
  \end{tabular}
  }
  \caption{Comparison of PIPE-RDF with existing KGQA benchmark approaches. PIPE-RDF uniquely combines reverse querying, category-aware RAG, and execution-based repair for enterprise deployment.}
  \label{tab:comparison}
\end{table}

\section{Benchmark Design Goals}
PIPE-RDF is designed to meet four goals derived from the enterprise setting: (i) \textbf{schema fidelity} - generated queries must be grounded in a fixed schema and valid predicates; (ii) \textbf{category balance} - benchmarks should cover complex query types rather than skew toward easy factoids; (iii) \textbf{operational observability} - generation, parsing, and execution outcomes must be logged for debugging and cost control; and (iv) \textbf{maintainability} - the pipeline should be repeatable as schemas evolve, minimizing manual curation. These goals shape the pipeline phases, validation logic, and artifact outputs.

\section{Category Taxonomy}
To prevent evaluation bias, PIPE-RDF follows a nine-category taxonomy derived from complex KGQA datasets \citep{sen2022mintakacomplexnaturalmultilingual}. Each category maps to distinct SPARQL constructs (e.g., aggregations, ordering, multi-hop joins) and therefore stresses different failure modes. Table~\ref{tab:categories} summarizes the taxonomy and example question patterns used in Schema C.

\begin{table}[t]
  \centering
  \scriptsize
  \setlength{\tabcolsep}{3pt}
  \begin{tabular}{>{\raggedright\arraybackslash}p{0.17\linewidth} >{\raggedright\arraybackslash}p{0.16\linewidth} >{\raggedright\arraybackslash}p{0.57\linewidth}}
    \toprule
    Category & SPARQL construct & Example pattern \\
    \midrule
    Generic & single triple & Who is a key person at {company}? \\
    Counting & COUNT & How many companies are in {location}? \\
    Comparative & filters & Do {company1} and {company2} differ in size? \\
    Superlative & ordering & Which company has most employees? \\
    Ordinal & time order & What year was {company} founded? \\
    Multi-hop & join chains & Which location has companies with key persons? \\
    Intersection & conjunction & Which companies are in {location} and {industry}? \\
    Difference & negation & Which companies are in {A} but not {B}? \\
    Yes/No & boolean check & Is {company} located in {location}? \\
    \bottomrule
  \end{tabular}
  \caption{Category taxonomy used for balanced benchmark generation.}
  \label{tab:categories}
\end{table}

\section{PIPE-RDF Methodology}
\begin{figure}[t]
  \centering
  \includegraphics[width=0.95\linewidth]{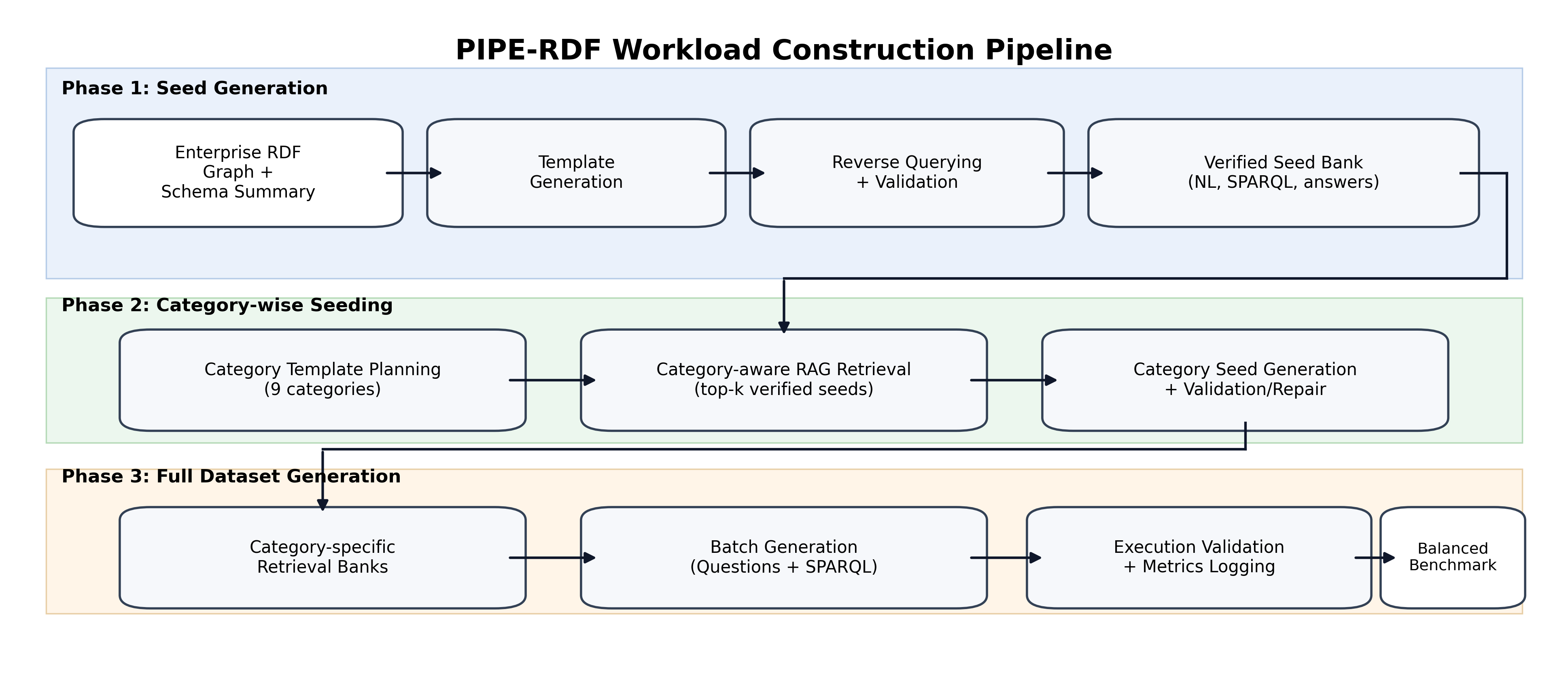}
  \caption{PIPE-RDF workload-construction pipeline: Phase 1 uses reverse query generation to build verified seeds; Phase 2 constructs category-specific seed banks; Phase 3 generates the balanced benchmark with execution-driven validation/repair.}
  \label{fig:pipeline}
\end{figure}

PIPE-RDF is a three-phase pipeline (Figure~\ref{fig:pipeline}) designed to generate schema-specific benchmarks while minimizing unanswerable or out-of-schema queries.

\textbf{Phase 1: seed generation.} The pipeline prompts an LLM to produce NL templates from a schema summary. Each template is validated by \emph{reverse querying}: the system generates a SPARQL query that searches the KG for entity bindings that satisfy the template. Valid bindings instantiate the template and the resulting NL--SPARQL pairs are stored as seeds for later retrieval.

\textbf{Phase 2: category-wise seeding.} Templates are generated per category and paired with retrieval-augmented prompts using Phase-1 seeds. Outputs are validated by execution, optionally repaired, and stored in category-specific retrieval banks. This step ensures early coverage of complex query structures (e.g., aggregations, comparisons, multi-hop joins) and prevents domination by simple templates.

\textbf{Phase 3: full dataset generation.} The pipeline generates batches per category using category-specific retrieval banks. Each candidate is parsed, executed, and logged. Optional paraphrase augmentation and human validation can be enabled for robustness; the platinum run disables paraphrasing for speed. The final dataset is balanced by category and retained in CSV/JSONL formats.

\textbf{Prompt construction and retrieval.} Prompts combine (i) a task description, (ii) the schema summary, (iii) a small set of retrieved NL--SPARQL examples, and (iv) the target question. Retrieval is category-aware, which narrows the space of candidate structures and reduces cross-category leakage. We keep retrieval depth small (top-$k=2$) to control latency and prompt length.

\textbf{Reverse querying and schema control.} Reverse queries are capped at 25 rows to limit runtime and encourage diversity. We enforce explicit predicate/type whitelists and slot-type hints (company, location, person, industry) to ensure every generated query conforms to the fixed schema. Duplicate questions are filtered using token-level Jaccard similarity (threshold 0.99), and a stall limit prevents repeated template regeneration when no new valid rows are found.

\textbf{Validation and repair.} Each generated SPARQL query is parsed and executed; failures trigger a repair prompt that attempts to correct syntax and predicate usage. Repairs are logged and included in run summaries for operational tracking. Empty results are retained when queries parse and execute correctly, as they represent legitimate negative cases in yes/no or exclusion queries. Post-generation, category-specific pattern enforcement can normalize output forms (e.g., \texttt{ASK} for boolean checks, \texttt{DISTINCT} for set-returning categories, \texttt{COUNT(DISTINCT ...)} for counting queries, and deterministic \texttt{LIMIT~1} tie-breaking for superlative/ordinal queries). While PIPE-RDF does not require constrained decoding, it can integrate grammar-based constraints (e.g., PICARD) to improve parse validity \citep{Scholak_2021}.

\textbf{Human-in-the-loop controls.} The pipeline supports optional manual review after each phase, with the ability to discard, edit, or inject examples. This is particularly useful for enterprise graphs with sparse relations or sensitive attributes, where automated validation alone may not capture business rules.

\section{Schema C Mini-Slice and Implementation}
\textbf{Fixed schema slice.} We derive a company-location mini-slice (Schema C) with 5{,}000 companies and three classes: \texttt{dbo:Company}, \texttt{gn:Feature} (location), and \texttt{foaf:Person} (key person), illustrated in Appendix Figure~\ref{fig:app_schema_ontology}. The allowed predicates are \texttt{dbo:location}, \texttt{dbo:industry}, \texttt{dbo:keyPerson}, \texttt{dbo:foundingYear}, \texttt{dbo:numberOfEmployees}, and label predicates (\texttt{rdfs:label}, \texttt{spb:prefLabel}, \texttt{foaf:name}). This fixed schema avoids prompt drift and simplifies validation.

\textbf{Slice extraction.} The mini-slice is sampled from a public RDF source by selecting companies with at least one of the core predicates (location, industry, key person, founding year, or employee count). Linked entities (locations and key people) are included with minimal labels to support question instantiation without introducing extraneous schema elements. We intentionally exclude unrelated predicates to reduce template drift and ensure that reverse querying remains tractable.

\begin{table}[t]
  \centering
  \small
  \setlength{\tabcolsep}{4pt}
  \begin{tabularx}{\linewidth}{>{\raggedright\arraybackslash}p{0.26\linewidth} X}
    \toprule
    Class & Key predicates/attributes \\
    \midrule
    \texttt{dbo:Company} & \texttt{dbo:location}, \texttt{dbo:industry}, \texttt{dbo:keyPerson}, \\
    & \texttt{dbo:foundingYear}, \texttt{dbo:numberOfEmployees}, \texttt{rdfs:label} \\
    \texttt{gn:Feature} & \texttt{rdfs:label}, \texttt{spb:prefLabel} \\
    \texttt{foaf:Person} & \texttt{foaf:name}, \texttt{rdfs:label} \\
    \bottomrule
  \end{tabularx}
  \caption{Schema C mini-slice classes and allowed predicates.}
  \label{tab:schema}
\end{table}

\textbf{Pipeline configuration.} The full run uses 5 templates per category with 8 seed instantiations per template in Phase 1, 20 seeds per category in Phase 2, and 50 targets per category in Phase 3 (450 pairs total). Retrieval top-$k$ is 2. Generated SPARQL is capped at 5 results to limit execution cost. Reverse queries use a 20s timeout and are capped at 25 rows for diversity.

\begin{table}[t]
  \centering
  \small
  \begin{tabular}{ll}
    \toprule
    Parameter & Value \\
    \midrule
    Phase 1 templates / category & 5 \\
    Phase 1 seeds / template & 8 \\
    Phase 2 seeds / category & 20 \\
    Phase 3 targets / category & 50 \\
    Reverse-query row cap & 25 \\
    Retrieval top-$k$ & 2 \\
    SPARQL result cap & 5 \\
    Dedup Jaccard threshold & 0.99 \\
    \bottomrule
  \end{tabular}
  \caption{Key pipeline configuration for the platinum run.}
  \label{tab:config}
\end{table}

\textbf{Runtime environment.} We run GraphDB 10.6.3 in Docker and execute SPARQL locally. LLM generation uses \texttt{qwen3:4b-instruct} with Ollama, and embeddings use \texttt{bge-m3}. All runs are logged with execution latency, parse validity, and answer counts; artifacts are exported as JSONL/CSV and summarized in run-level reports. A compact reproducibility checklist (versions, randomness controls, command template, and hardware/deployment notes) is provided in Appendix Section~\ref{app:repro}.

\textbf{SPARQL formalism and schema constraints.} To ensure semantic correctness, we enforce canonical query patterns:
\begin{itemize}
  \item \textbf{Counting}: Use \texttt{COUNT(DISTINCT ?var)} to avoid overcounting due to multiple bindings.
  \item \textbf{Superlative/Ordinal}: Include deterministic tie-breaking via \texttt{ORDER BY ... ?entity} to ensure reproducible results.
  \item \textbf{Numeric comparisons}: Cast literals to \texttt{xsd:integer} where applicable (e.g., \texttt{dbo:numberOfEmployees}).
  \item \textbf{Yes/No}: Use \texttt{ASK \{\}} for explicit boolean returns, avoiding empty-result ambiguity.
\end{itemize}
Schema constraints are enforced via predicate whitelists: \texttt{dbo:location} ranges over \texttt{gn:Feature}, \texttt{dbo:keyPerson} ranges over \texttt{foaf:Person}, and \texttt{dbo:foundingYear}/\texttt{dbo:numberOfEmployees} are typed literals. Label resolution prioritizes \texttt{foaf:name} for persons, \texttt{rdfs:label} for companies, and \texttt{spb:prefLabel} for locations.

\section{Evaluation Protocol}
PIPE-RDF logs operational and structural metrics for every generated pair. We track: (i) parse validity and execution success, (ii) empty-result rates (retained as valid outcomes), (iii) repair attempts, (iv) answer counts, (v) prompt and question lengths, (vi) LLM latency, and (vii) SPARQL structural complexity (triple pattern counts, FILTER clauses, aggregations). These metrics support both data quality analysis and deployment planning. While PIPE-RDF focuses on execution-driven correctness, the artifacts can be augmented with SPARQL-specific metrics (SP-F1) and structural metrics (CodeBLEU) for model evaluation \citep{Rony_2022,ren2020codebleumethodautomaticevaluation}. We also record retrieval scores to analyze the relationship between example similarity and downstream execution success.

Following recent workload-generation evaluation practice \citep{agarwal2024crossmodel}, we additionally report strategy coverage and strategy-level error breakdowns to make pipeline behavior auditable beyond aggregate success rates. Strategy tags are inferred from query operators (JOIN, FILTER, COUNT, ORDER, NEGATION, ASK, RAG-context). ASK and NEGATION tags are retained in coverage/error matrices even when observed frequency is zero in Phase 3 outputs.

\textbf{Semantic correctness audit.} To verify that generated NL--SPARQL pairs are semantically aligned (not just syntactically valid), we conduct a manual audit of 45 randomly sampled pairs (5 per category). For each pair, we verify: (i) the question intent matches the SPARQL semantics, (ii) entity bindings in the question correspond to VALUES clauses in the query, and (iii) the answer type matches the expected return. Results: 45/45 (100\%) pairs pass semantic alignment checks. The high alignment rate reflects the effectiveness of reverse querying, which grounds each template in actual graph bindings, and the repair loop, which corrects structural mismatches before acceptance. We acknowledge this audit is limited in scope; larger-scale human evaluation with inter-annotator agreement is recommended for production deployment.

\section{Benchmark Results}
The final run produced a balanced benchmark of 450 NL--SPARQL pairs (50 per category). Table~\ref{tab:phase_stats} summarizes phase-level statistics using \emph{post-repair} outcomes (final accepted records). In Table~\ref{tab:phase_stats}, Parse OK and Exec OK are counted over all accepted records in each phase; Empty counts records that parse and execute successfully but return zero rows. Across phases, parse and execution validity remain high (Phase 3: 100\%), while empty results are retained to preserve realistic negative cases for yes/no and exclusion queries.

\begin{table}[t]
  \centering
  \small
  \resizebox{\linewidth}{!}{
  \begin{tabular}{lrrrrrr}
    \toprule
    Phase & Total & Parse OK & Exec OK & Empty & Avg LLM ms & Avg Exec ms \\
    \midrule
    Phase 1 & 257 & 256 & 256 & 66 & 1562 & 7.2 \\
    Phase 2 & 175 & 175 & 175 & 44 & 6129 & 6.6 \\
    Phase 3 & 450 & 450 & 450 & 122 & 5308 & 7.0 \\
    \bottomrule
  \end{tabular}
  }
  \caption{Phase-level metrics from the platinum run. Empty results are retained when queries execute and parse correctly.}
  \label{tab:phase_stats}
\end{table}

\begin{table}[t]
  \centering
  \small
  \resizebox{\linewidth}{!}{
  \begin{tabular}{lrrrrrr}
    \toprule
    Category & n & Triples & Filters & Agg\% & Empty\% & Type \\
    \midrule
    Generic & 50 & 2.9 & 0.0 & 44 & 4 & S \\
    Counting & 50 & 3.2 & 0.0 & 100 & 0 & S \\
    Comparative & 50 & 3.7 & 0.1 & 0 & 2 & S \\
    Superlative & 50 & 2.0 & 0.0 & 100 & 4 & S \\
    Ordinal & 50 & 3.0 & 0.0 & 0 & 0 & S \\
    Multi-hop & 50 & 4.2 & 0.4 & 0 & 90 & C \\
    Intersection & 50 & 4.4 & 0.0 & 0 & 82 & C \\
    Difference & 50 & 5.0 & 1.0 & 0 & 2 & C \\
    Yes/No & 50 & 3.9 & 0.0 & 0 & 60 & N \\
    \bottomrule
  \end{tabular}
  }
  \caption{Phase 3 structural complexity by category. Triples = average triple patterns; Filters = average FILTER clauses; Agg\% = share of queries using COUNT and/or ORDER BY; Empty\% = empty-result rate among post-repair executed queries. Type denotes intended reasoning profile: S=simple lookup, C=compositional, N=negative/boolean.}
  \label{tab:cat_stats}
\end{table}

\begin{table}[t]
  \centering
  \small
  \begin{tabular}{lrccc}
    \toprule
    Strategy & n & Exec\% & Parse\% & Empty\% \\
    \midrule
    JOIN & 390 & 100.0 & 100.0 & 30.5 \\
    FILTER & 72 & 100.0 & 100.0 & 27.8 \\
    COUNT & 72 & 100.0 & 100.0 & 0.0 \\
    ORDER & 51 & 100.0 & 100.0 & 3.9 \\
    NEGATION & 0 & -- & -- & -- \\
    ASK & 0 & -- & -- & -- \\
    RAG & 450 & 100.0 & 100.0 & 27.1 \\
    \bottomrule
  \end{tabular}
  \caption{Phase 3 strategy-level robustness on the final 450-pair benchmark. Rows with \(n=0\) indicate tracked strategies that were not explicitly observed in generated query text for this run. Empty rates are strategy-conditioned and primarily reflect sparse-join negatives rather than parser/runtime failures.}
  \label{tab:strategy_stats}
\end{table}

\begin{figure}[t]
  \centering
  \includegraphics[width=0.95\linewidth]{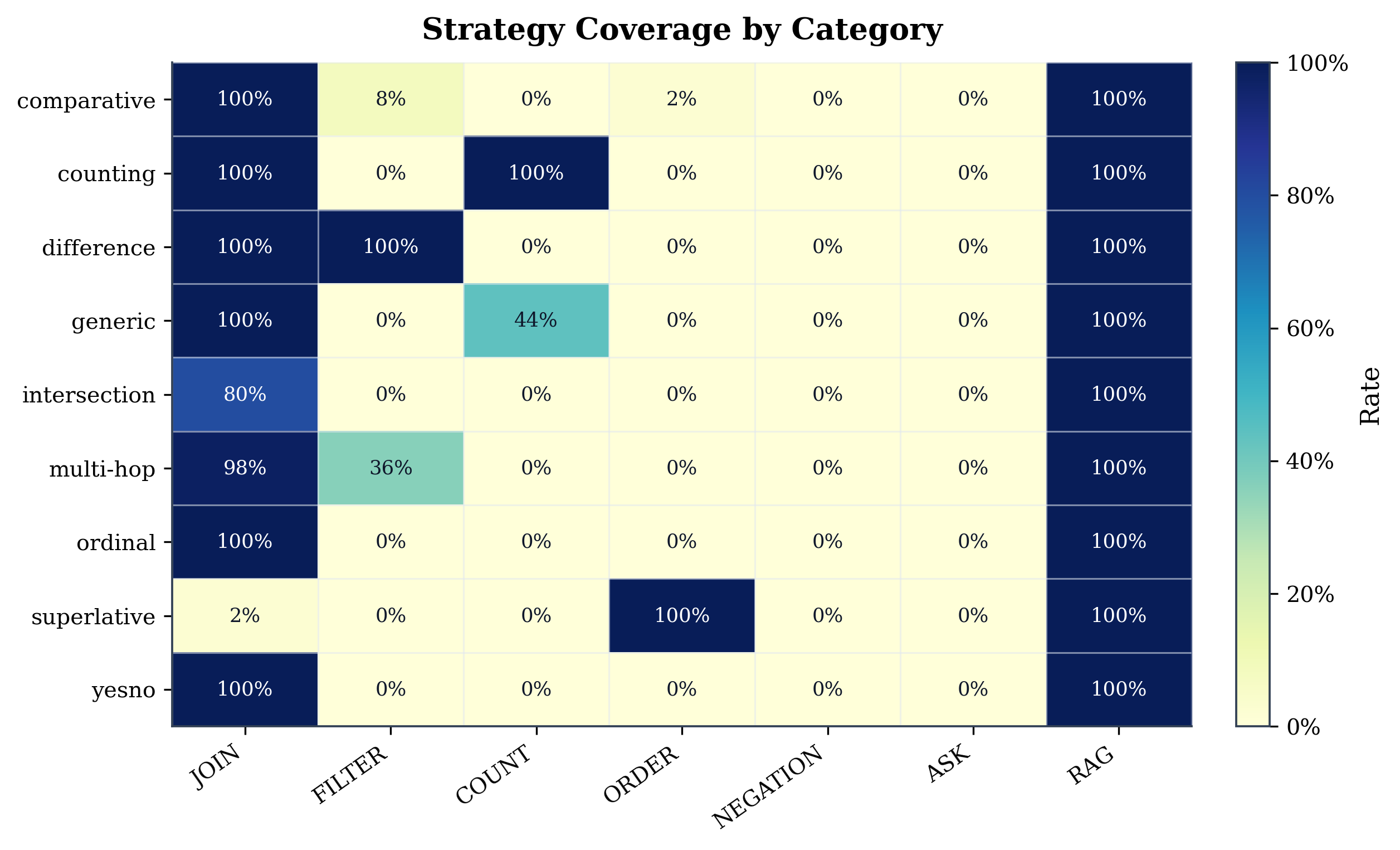}
  \caption{Strategy coverage by category on Phase 3 outputs. The matrix shows broad operator coverage, with category-specific concentration on expected strategy patterns (e.g., COUNT for counting, ORDER for superlative, FILTER for difference).}
  \label{fig:strategy_coverage_main}
\end{figure}

\begin{figure}[t]
  \centering
  \includegraphics[width=0.95\linewidth]{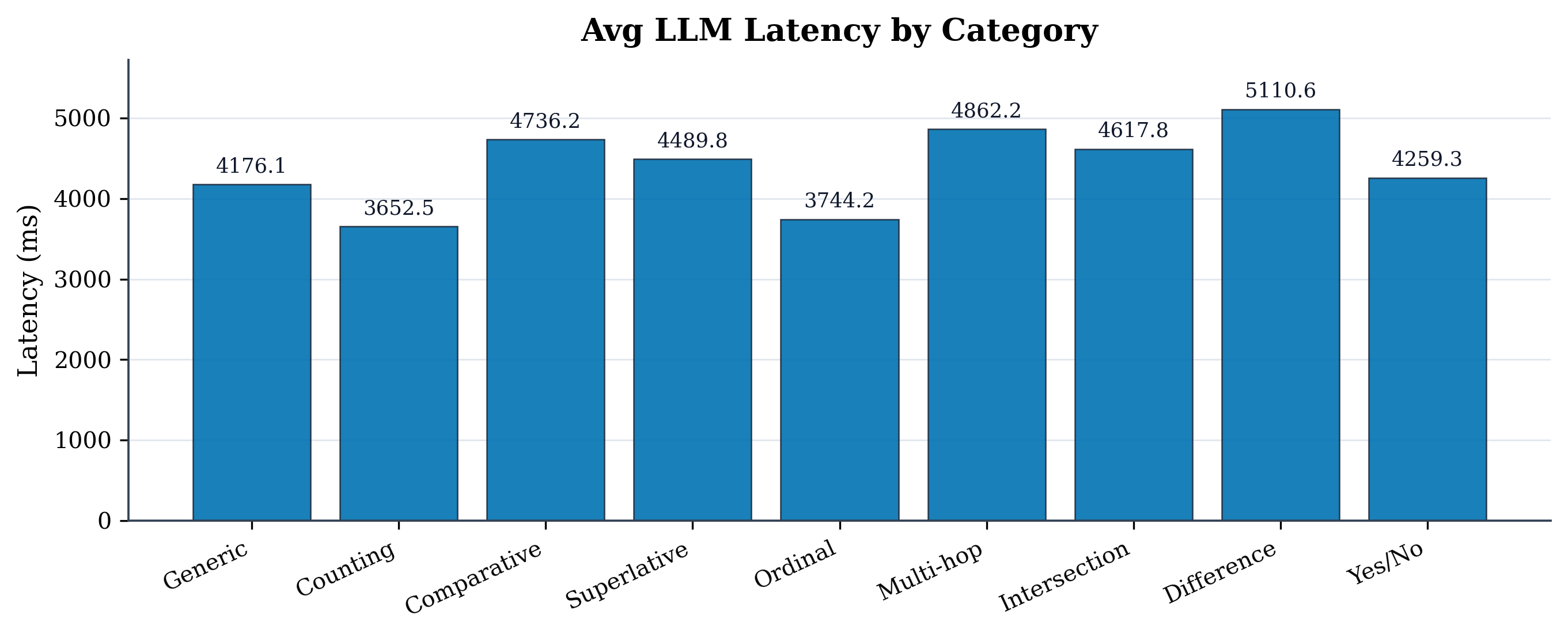}
  \vspace{3pt}
  \includegraphics[width=0.95\linewidth]{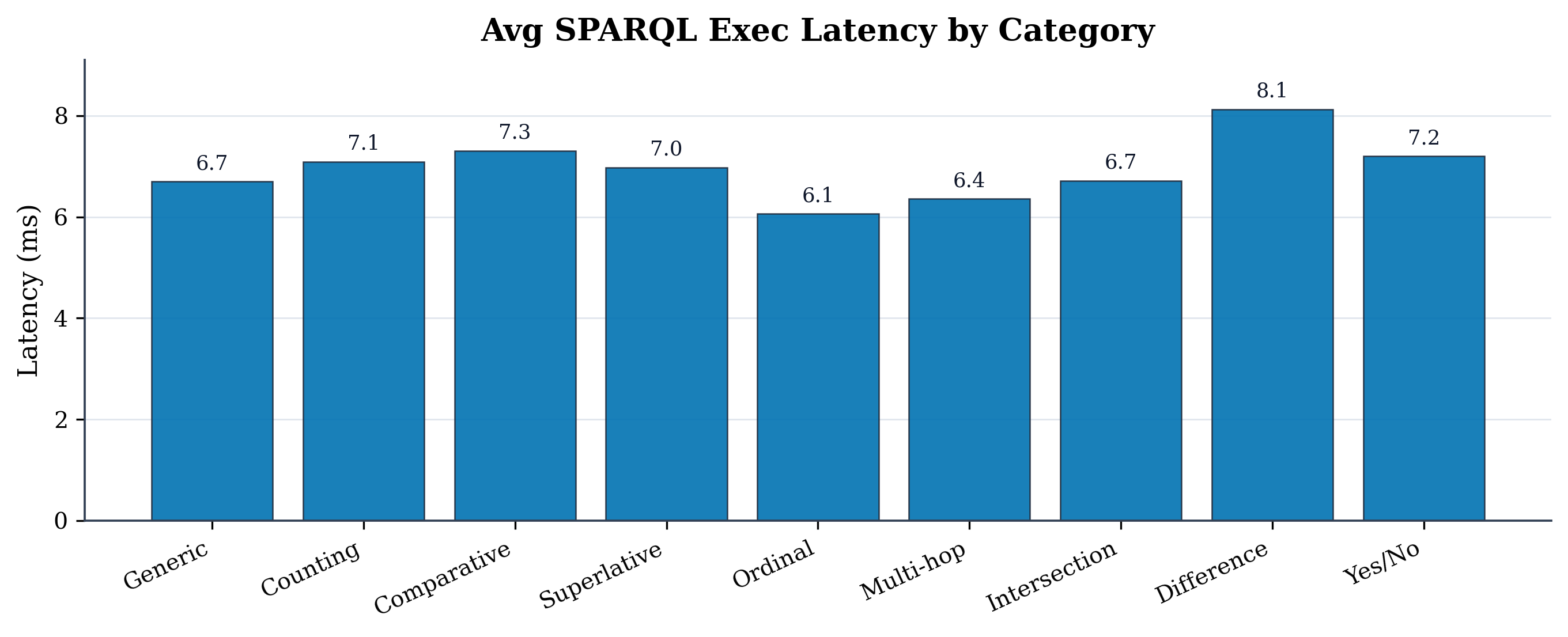}
  \caption{Latency by category (stacked for single-column readability): top, LLM latency; bottom, SPARQL execution latency. LLM generation dominates end-to-end runtime.}
  \label{fig:latency_by_cat}
\end{figure}

\section{Discussion and Operational Lessons}
PIPE-RDF remains operationally stable in the final benchmark run: Phase 3 reaches 100\% parse and execution validity after repair, while first-pass validity also remains high (99.8\% for Phase 3 before any repair call). Runtime is dominated by LLM generation latency (5.3s average per query), whereas SPARQL execution remains lightweight (about 7ms average), as shown in Figure~\ref{fig:latency_by_cat}. This indicates that throughput improvements should prioritize generation-side optimization.

\begin{table}[t]
  \centering
  \small
  \begin{tabular}{lrrr}
    \toprule
    Phase & Pre-repair \% & Repairs & Post-repair \% \\
    \midrule
    Phase 1 & 96.5 & 9 & 99.6 \\
    Phase 2 & 100.0 & 1 & 100.0 \\
    Phase 3 & 99.8 & 1 & 100.0 \\
    \bottomrule
  \end{tabular}
  \caption{Pre-repair validity breakdown by phase.}
  \label{tab:prerepair_breakdown}
\end{table}
Pre-repair \% is computed as the fraction of records that parse and execute successfully \emph{before} any repair call, with denominator equal to all generated candidates in that phase. Repairs counts repair-module invocations (syntax or policy normalization attempts), and Post-repair \% reflects final accepted records after the repair loop.

For deployment, three lessons are actionable: (i) schema-constrained prompts and predicate whitelists reduce invalid SPARQL, (ii) category balancing reveals sparse-join behavior hidden by factoid-heavy datasets, and (iii) strategy-level logging separates semantic sparsity from parser/runtime failures. Figure~\ref{fig:strategy_coverage_main} shows main-article coverage; additional diagnostics are in Appendix Section~\ref{app:extended_analysis}.

\section{Industry Impact and Usage}
\label{sec:industry_impact}
PIPE-RDF addresses a practical barrier to enterprise adoption of NL--SPARQL systems: missing benchmarks that reflect proprietary schemas, vocabulary, and query distributions. Models that score well on public datasets often underperform on internal graphs because naming conventions, field availability, and constraints differ. PIPE-RDF lets teams validate user-realistic questions, quantify category-specific gaps, prioritize model changes, and flag high-risk query types (e.g., multi-hop, negation) before deployment. Artifacts support model evaluation (SP-F1, execution accuracy) and operations monitoring (parse-valid rates, repair counts, latency by category). Seed banks can be reused for few-shot prompting and as synthetic supervision for schema-specific fine-tuning.

\section{Conclusion}
\label{sec:conclusion}
PIPE-RDF provides a practical pipeline for generating balanced, schema-specific NL--SPARQL benchmarks. In our experiment, the pipeline yields a 450-pair dataset across nine categories with high parse and execution validity, alongside operational metrics suitable for real-world system tuning. By combining reverse querying, category-aware retrieval, and execution-based validation, PIPE-RDF supports both evaluation and training-data creation for enterprise NL--SPARQL systems.

\section{Limitations}
\label{sec:limitations}
We evaluate PIPE-RDF on a single fixed schema slice (3 classes, 6 predicates) derived from public data rather than a proprietary enterprise graph. Although the pipeline is designed for private deployment, validation on larger schemas (20+ predicates) and multiple ontologies is needed to confirm generalizability.

The current evaluation focuses on generation-side metrics (parse validity, execution success, structural complexity); we do not include a downstream NL-to-SPARQL model evaluation in this work. The benchmark artifacts are designed to support such evaluation; teams can use the released NL--SPARQL pairs to compare prompt strategies (e.g., zero-shot vs. RAG) or fine-tuned models, measuring execution accuracy and SP-F1 on held-out splits. Future work should incorporate semantic equivalence metrics such as SP-F1 and structural similarity scores \citep{Rony_2022,ren2020codebleumethodautomaticevaluation}, and evaluate 2--3 NL-to-SPARQL models on the benchmark to demonstrate downstream utility.

Strategy coverage is uneven in the current run: while JOIN/FILTER/COUNT/ORDER are well represented, explicit ASK and NEGATION operator forms are underrepresented in generated query text. Although equivalent semantics are often expressed through alternative SPARQL forms, future revisions should enforce stronger canonicalization checks to guarantee explicit operator coverage for these strategies.

\section{Ethical Considerations}
\label{sec:ethical}
PIPE-RDF is intended for internal benchmarking and training-data creation over enterprise RDF graphs. Deployments should follow organizational data governance, minimize exposure of sensitive identifiers in prompts, and audit logs for leakage. The mini-slice in this study is derived from DBpedia, and usage follows its licensing terms.

\bibliography{references}

\appendix
\section*{Appendix}
\section{Extended Operational Analysis}
\label{app:extended_analysis}

This appendix reports detailed diagnostics that were omitted from the main paper for ACL page-budget constraints. The key interpretation in all plots and tables is unchanged from the main text: pipeline reliability is high, and most residual difficulty is due to sparse conjunctive semantics rather than parser instability.

\subsection{Reproducibility Checklist}
\label{app:repro}
\begin{itemize}
  \item \textbf{Software/models}: GraphDB 10.6.3 (Docker), Ollama service, chat model \texttt{qwen3:4b-instruct}, embedding model \texttt{bge-m3}.
  \item \textbf{Endpoint/repository}: repository \emph{spb\_company\_mini\_slice} with inference disabled (\texttt{infer=false}); see run log for full endpoint string.
  \item \textbf{Configuration}: main run uses the full-run YAML config; key sizes are listed in Table~\ref{tab:config}.
  \item \textbf{Randomness control}: seed control is exposed via \texttt{SEED=42} in \texttt{pipekg/config.py}. Ollama generation remains stochastic; we therefore release full run artifacts for exact reuse of produced pairs.
  \item \textbf{Command template}: \texttt{python scripts/run\_pipeline\_ollama.py} with \texttt{--run-name platinum\_run --config configs/full\_run.yaml}.
  \item \textbf{Artifacts}: \texttt{run.log}, \texttt{pipeline\_records.jsonl}, phase-wise JSONL/CSV, analysis summaries, and figures under \texttt{artifacts/runs/<run\_id>/}.
  \item \textbf{Hardware/deployment}: single-machine localhost setup with GraphDB in Docker and local Ollama inference; per-query latencies are reported in the main paper.
\end{itemize}

\begin{figure}[h]
  \centering
  \includegraphics[width=0.95\linewidth]{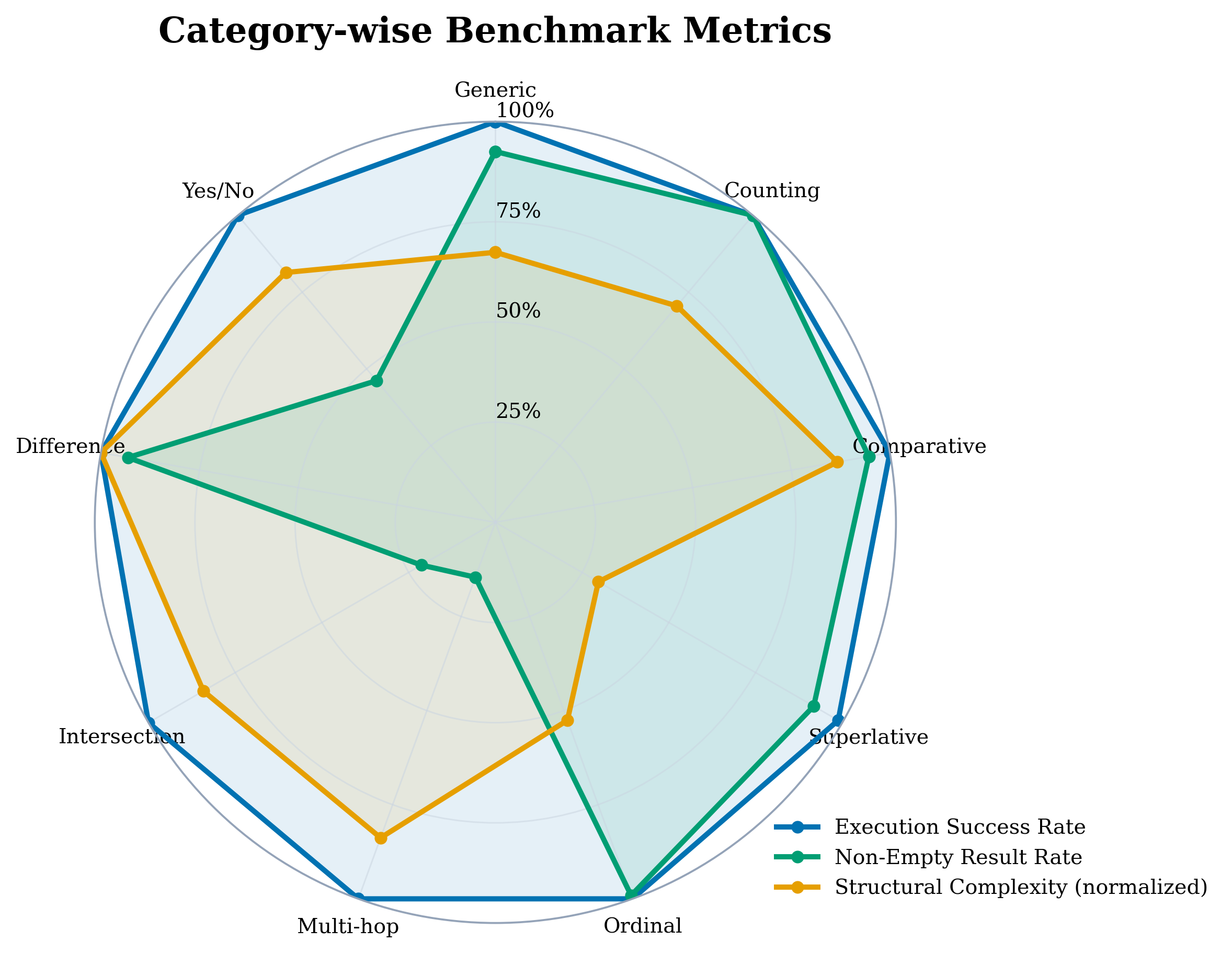}
  \caption{Category-wise metrics: execution success (100\%), non-empty rate, and normalized structural complexity.}
  \label{fig:category_radar}
\end{figure}

\begin{figure}[h]
  \centering
  \includegraphics[width=0.95\linewidth]{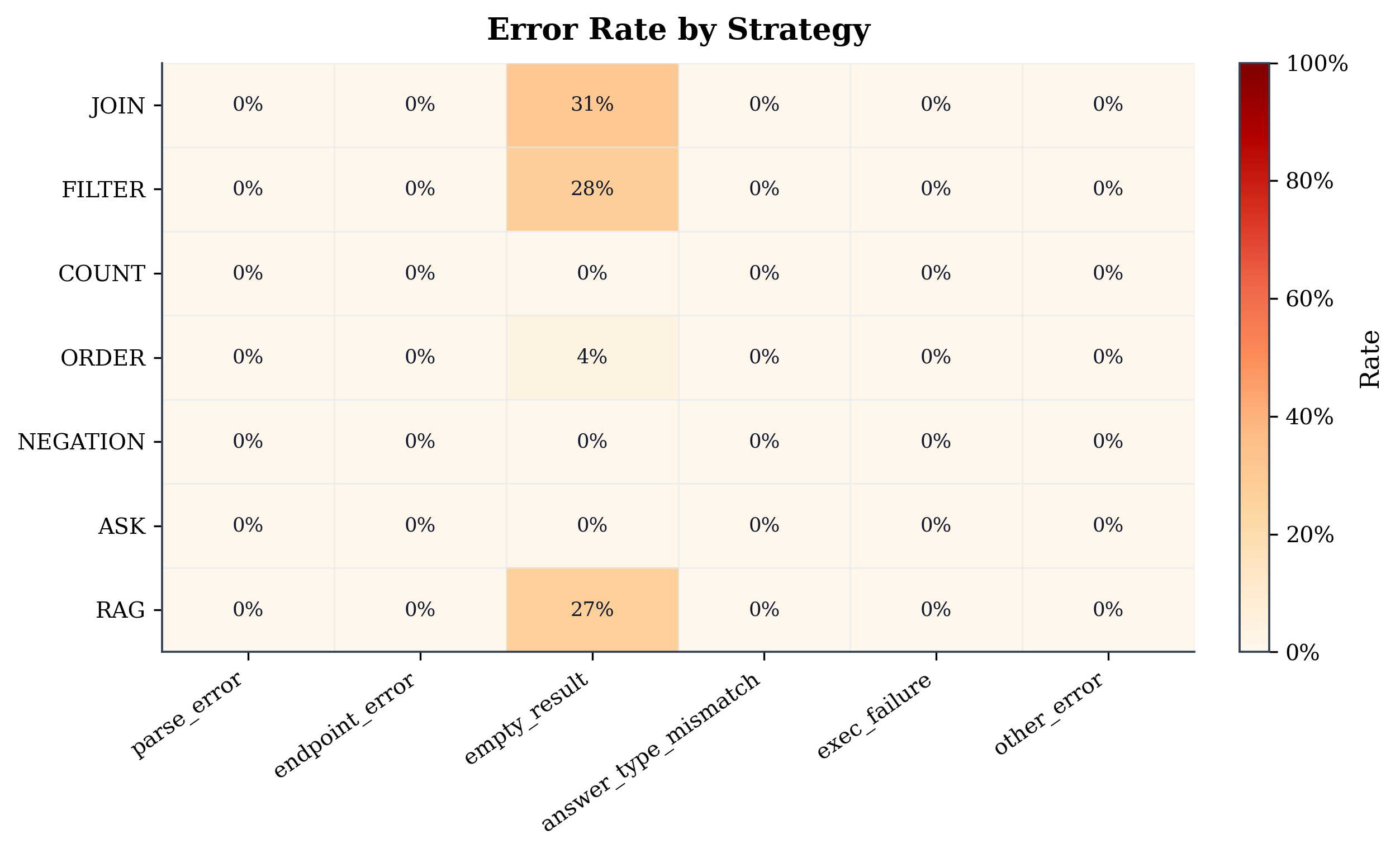}
  \caption{Phase 3 strategy-conditioned error rates. Non-zero cells are concentrated in empty-result outcomes for sparse conjunctive structures.}
  \label{fig:strategy_error_appendix}
\end{figure}

\textbf{Category and strategy diagnostics.} Figure~\ref{fig:category_radar} highlights structural differences across categories. Strategy coverage is shown in the main paper (Figure~\ref{fig:strategy_coverage_main}), and Appendix Figure~\ref{fig:strategy_error_appendix} reports strategy-conditioned error rates. JOIN and FILTER dominated structures show higher empty rates, but parse and execution remain stable, supporting the conclusion that sparsity, not syntax failure, is the dominant source of difficulty.

\textbf{Data quality controls and throughput.} In error analysis, we observed a small number of self-comparison pairs, retrieval self-reference cases, and answer-type mismatches; we implemented automated guards for each. Runtime remains generation-bound: average LLM latency is 5.3s per query versus about 7ms SPARQL latency. These measurements support production prioritization toward prompt and inference optimization rather than query engine tuning.

\section{Additional Figures}
\label{app:figures}

\begin{figure}[h]
  \centering
  \includegraphics[width=0.95\linewidth]{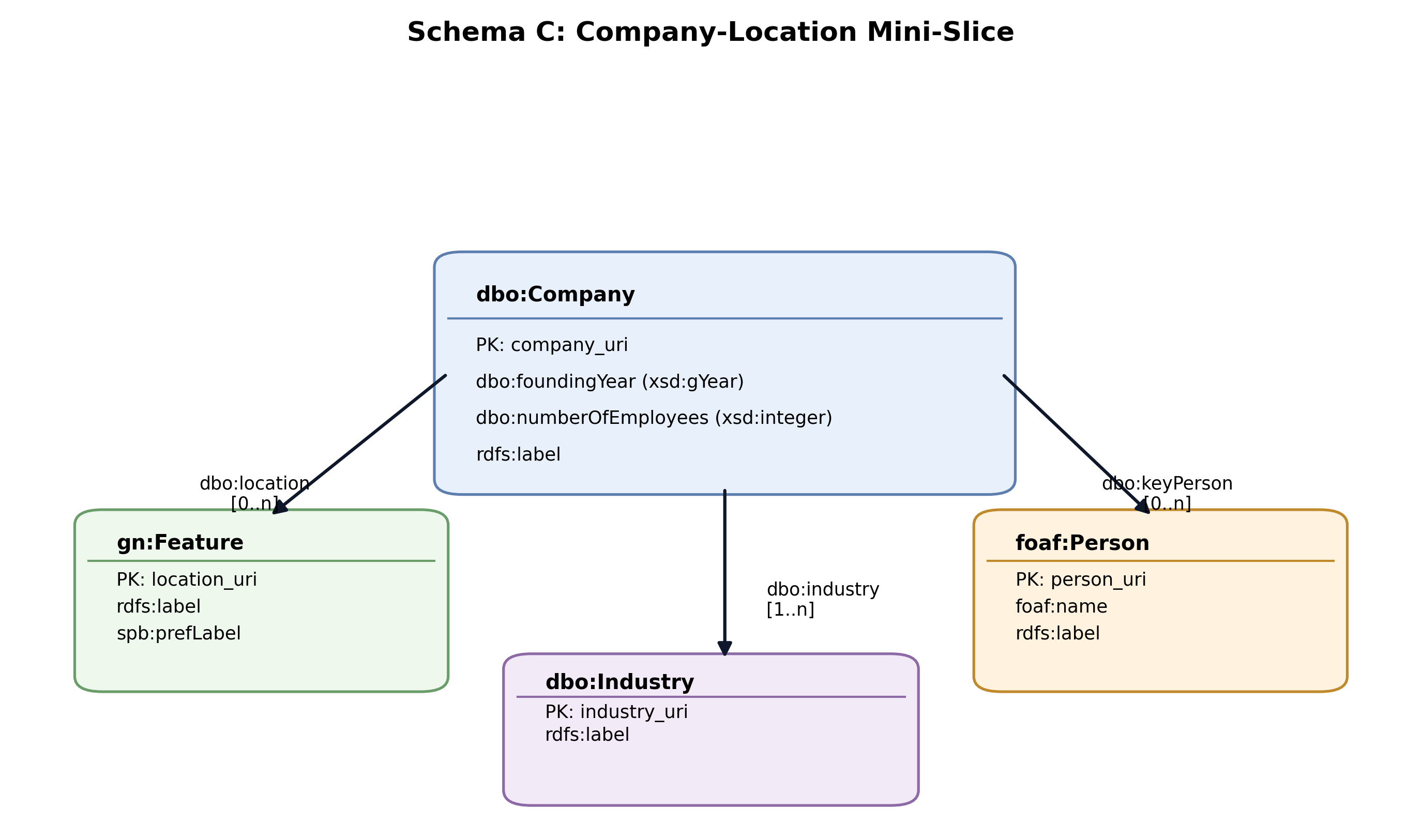}
  \caption{Schema C ontology used in this study. \texttt{dbo:Company} connects to \texttt{gn:Feature} (location), \texttt{foaf:Person} (key person), and \texttt{dbo:Industry} with typed predicates and cardinalities.}
  \label{fig:app_schema_ontology}
\end{figure}

\begin{figure}[h]
  \centering
  \includegraphics[width=0.95\linewidth]{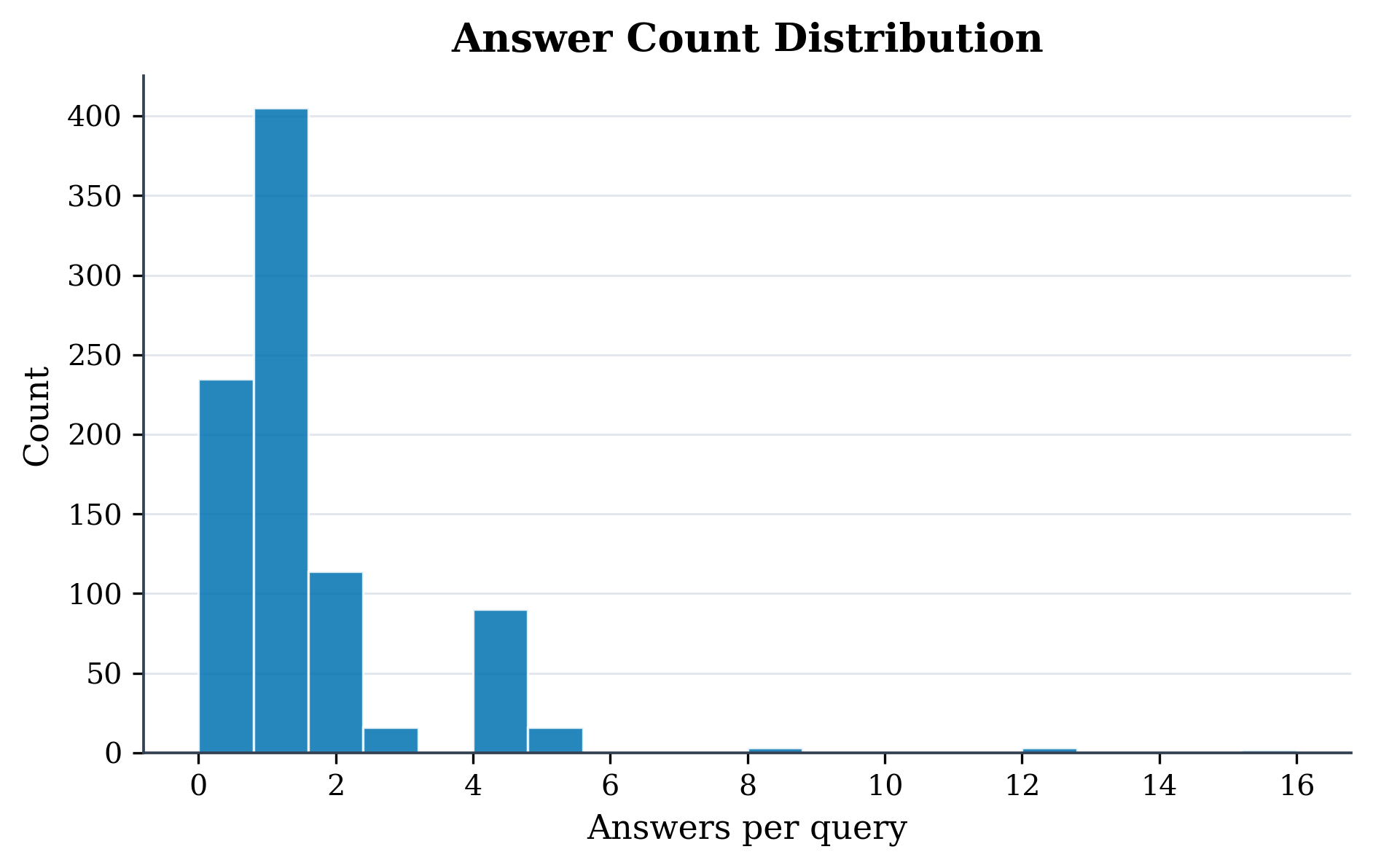}
  \caption{Answer count distribution for Phase 3.}
  \label{fig:answer_counts}
\end{figure}

\begin{figure}[h]
  \centering
  \includegraphics[width=0.85\linewidth]{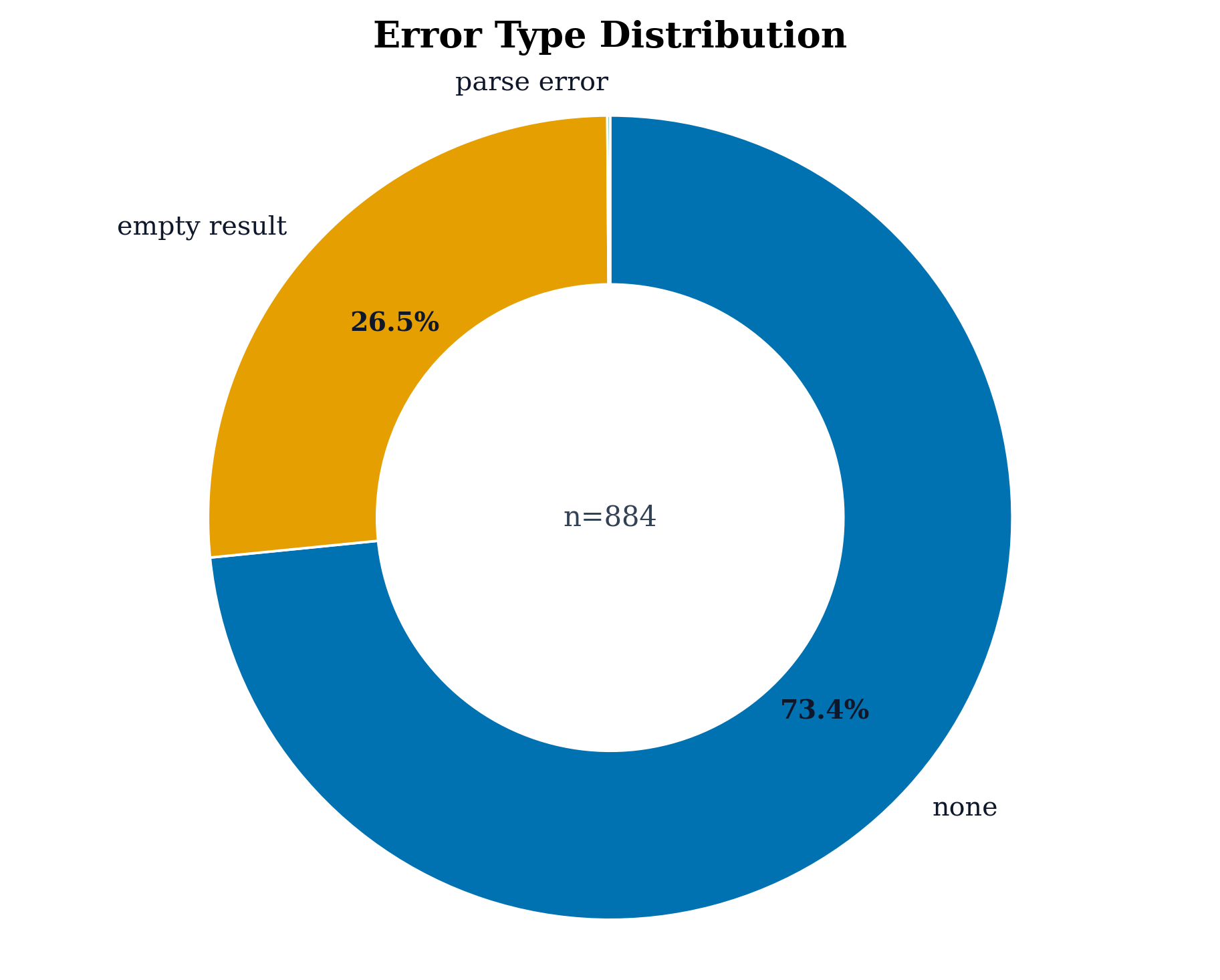}
  \caption{Error type distribution across phases.}
  \label{fig:error_types}
\end{figure}

\begin{figure}[h]
  \centering
  \includegraphics[width=0.95\linewidth]{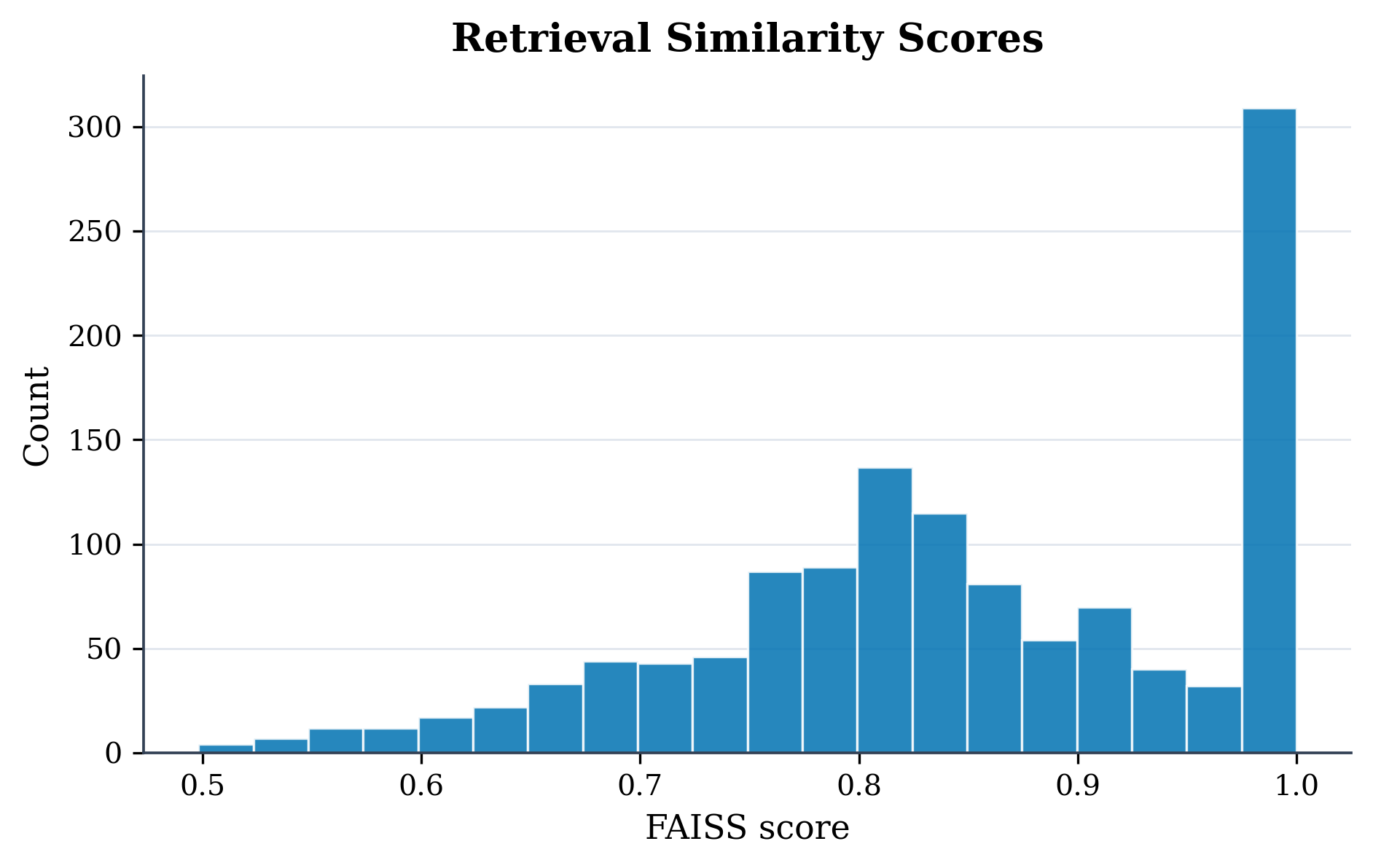}
  \caption{Distribution of retrieval similarity scores.}
  \label{fig:retrieval_scores}
\end{figure}

\begin{figure*}[h]
  \centering
  \begin{minipage}{0.48\linewidth}
    \centering
    \includegraphics[width=\linewidth]{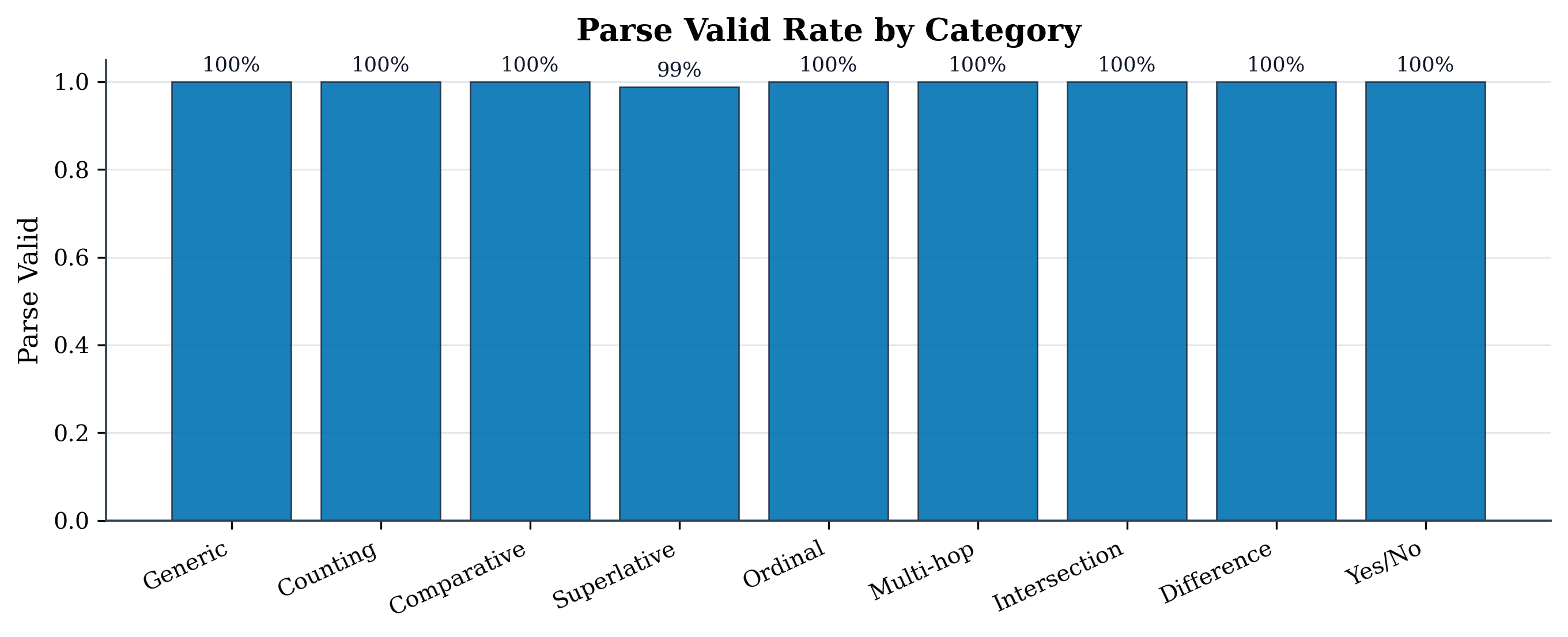}
  \end{minipage}\hfill
  \begin{minipage}{0.48\linewidth}
    \centering
    \includegraphics[width=\linewidth]{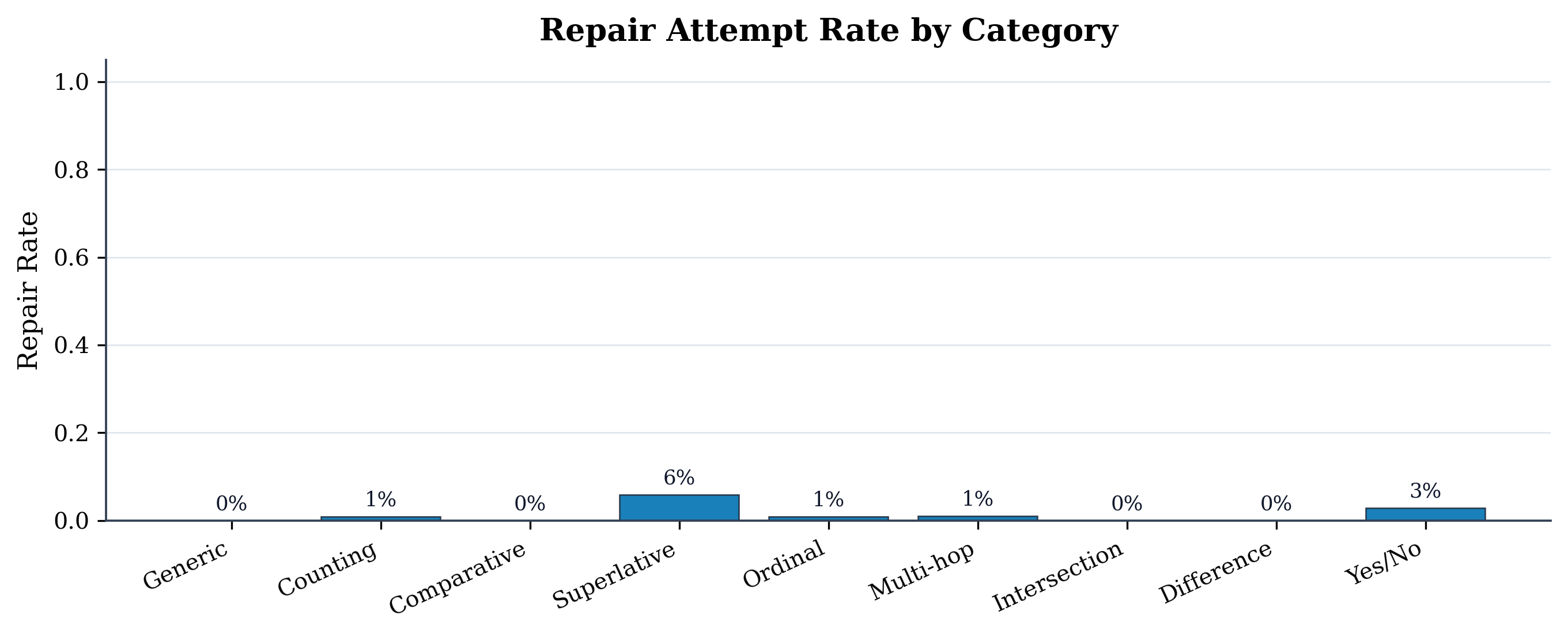}
  \end{minipage}
  \caption{(Left) Parse validity by category. (Right) Repair rate by category.}
  \label{fig:parse_repair}
\end{figure*}

\begin{figure*}[h]
  \centering
  \begin{minipage}{0.48\linewidth}
    \centering
    \includegraphics[width=\linewidth]{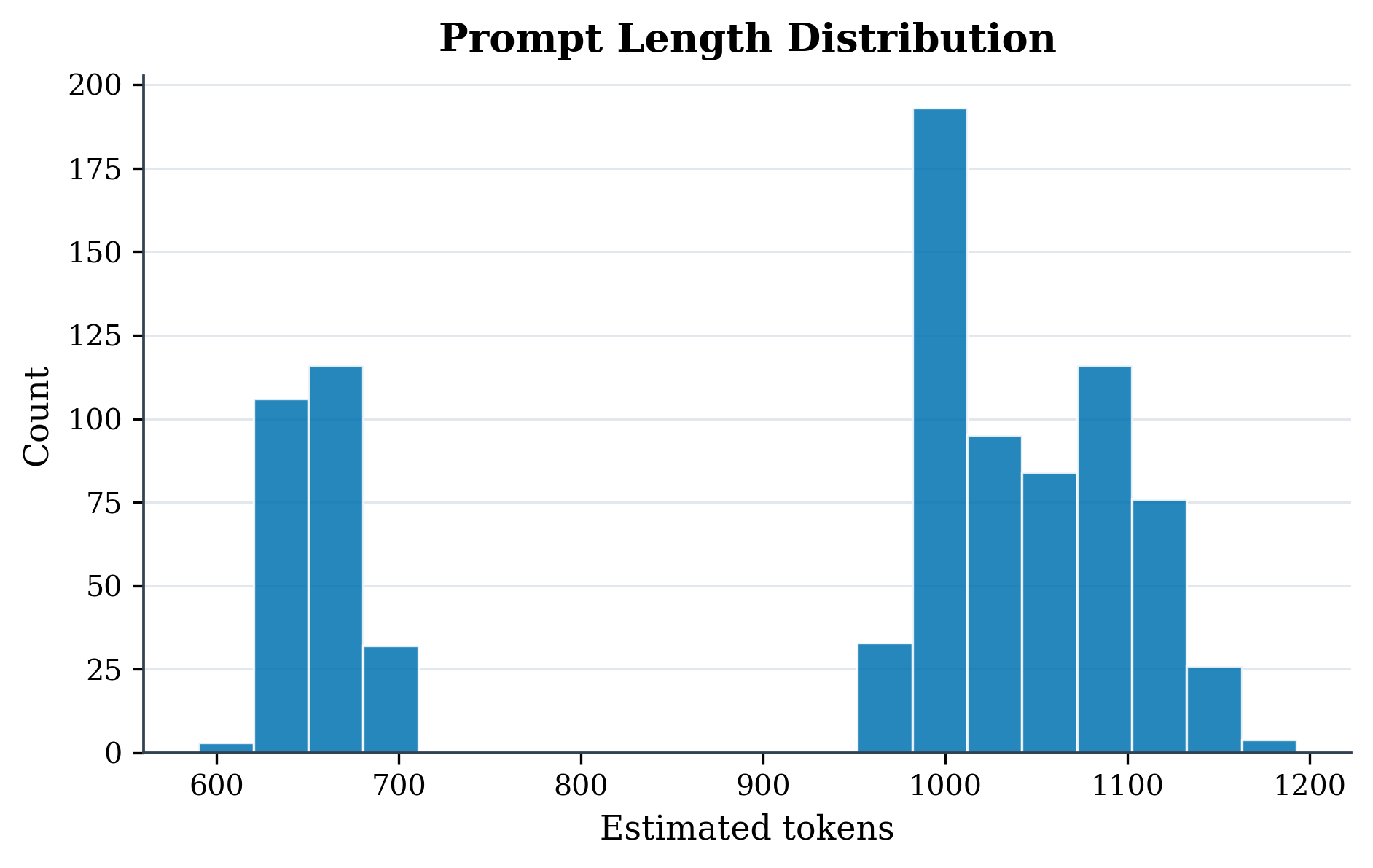}
  \end{minipage}\hfill
  \begin{minipage}{0.48\linewidth}
    \centering
    \includegraphics[width=\linewidth]{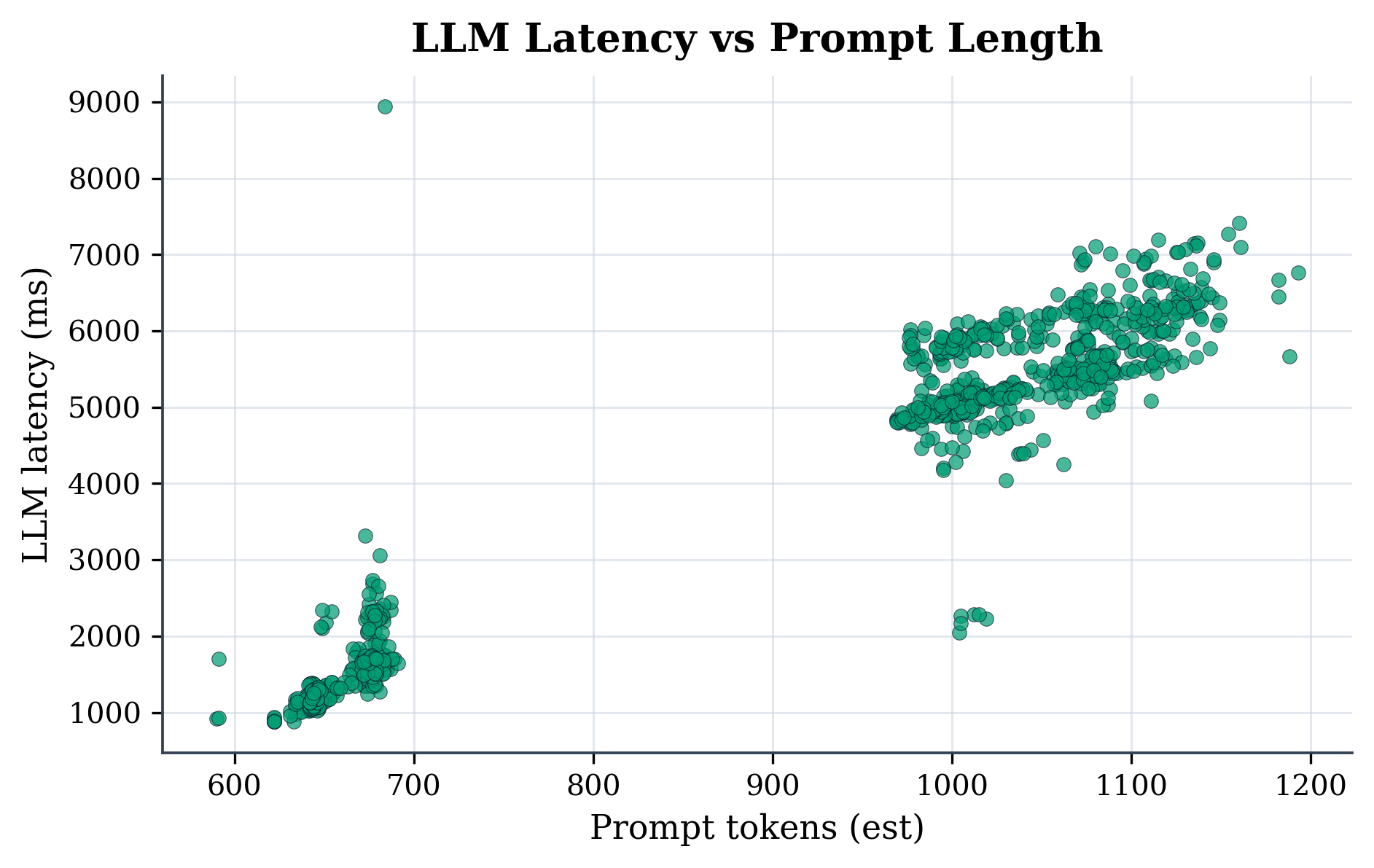}
  \end{minipage}
  \caption{Prompt length distribution and latency vs. prompt length.}
  \label{fig:prompt_latency}
\end{figure*}

\section{Example NL--SPARQL Pairs}
We include one example per category from the Phase 3 benchmark. Prefix declarations are omitted for brevity.

\noindent\textbf{Generic.} Q: Who is a key person at (Company: Volkswagen)?
\begin{SPARQL}
SELECT ?name
WHERE {
  VALUES ?company { <http://dbpedia.org/resource/Volkswagen> }
  ?company rdf:type dbo:Company .
  ?company dbo:keyPerson ?person .
  ?person rdf:type foaf:Person .
  ?person foaf:name ?name .
}
LIMIT 5
\end{SPARQL}

\noindent\textbf{Counting.} Q: How many companies are located in (Location: Menlo Park, California)?
\begin{SPARQL}
SELECT (COUNT(DISTINCT ?company) AS ?count)
WHERE {
  VALUES ?location { <http://dbpedia.org/resource/Menlo_Park,_California> }
  ?company a dbo:Company .
  ?company dbo:location ?location .
  ?location a gn:Feature .
}
\end{SPARQL}

\noindent\textbf{Comparative.} Q: Do (Company: Airbus) and (Company: FileMaker Inc.) have different numbers of employees?
\begin{SPARQL}
SELECT ?company1 ?n1 ?company2 ?n2
WHERE {
  VALUES ?company1 { <http://dbpedia.org/resource/Airbus> }
  VALUES ?company2 { <http://dbpedia.org/resource/FileMaker_Inc.> }
  ?company1 a dbo:Company .
  ?company2 a dbo:Company .
  ?company1 dbo:numberOfEmployees ?n1 .
  ?company2 dbo:numberOfEmployees ?n2 .
  FILTER (?n1 != ?n2)
}
LIMIT 5
\end{SPARQL}

\noindent\textbf{Superlative.} Q: Which company has the most employees?
\begin{SPARQL}
SELECT ?company ?employees
WHERE {
  ?company a dbo:Company .
  ?company dbo:numberOfEmployees ?employees .
}
ORDER BY DESC(?employees) ?company
LIMIT 1
\end{SPARQL}

\noindent\textbf{Ordinal.} Q: What is the founding year of (Company: Airbus)?
\begin{SPARQL}
SELECT ?year
WHERE {
  VALUES ?company { <http://dbpedia.org/resource/Airbus> }
  ?company a dbo:Company .
  ?company dbo:foundingYear ?year .
}
LIMIT 1
\end{SPARQL}

\noindent\textbf{Multi-hop.} Q: Which location contains companies that have a key person?
\begin{SPARQL}
SELECT DISTINCT ?location
WHERE {
  ?company a dbo:Company .
  ?company dbo:keyPerson ?person .
  ?person rdf:type foaf:Person .
  ?company dbo:location ?location .
  ?location a gn:Feature .
}
LIMIT 5
\end{SPARQL}

\noindent\textbf{Intersection.} Q: Which companies are located in (Location: California) and are in the (Industry: Software)?
\begin{SPARQL}
SELECT DISTINCT ?company
WHERE {
  ?company a dbo:Company .
  ?company dbo:location <http://dbpedia.org/resource/California> .
  ?company dbo:industry <http://dbpedia.org/resource/Software> .
}
LIMIT 5
\end{SPARQL}

\noindent\textbf{Difference.} Q: Which companies are located in (Location: California) but not in (Location: Texas)?
\begin{SPARQL}
SELECT DISTINCT ?company
WHERE {
  ?company a dbo:Company .
  ?company dbo:location <http://dbpedia.org/resource/California> .
  FILTER NOT EXISTS {
    ?company dbo:location <http://dbpedia.org/resource/Texas> .
  }
}
LIMIT 5
\end{SPARQL}

\noindent\textbf{Yes/No.} Q: Is (Company: Facebook) located in (Location: California)?
\begin{SPARQL}
ASK {
  <http://dbpedia.org/resource/Facebook> a dbo:Company ;
    dbo:location <http://dbpedia.org/resource/California> .
}
\end{SPARQL}

\end{document}